\documentclass[twoside,a4paper]{article}
\usepackage{amsmath,graphicx,amssymb,fancyhdr,amsthm,textcomp}  

\newtheorem{thm}{Theorem}[section]

\newtheorem{lem}[thm]{Lemma}
\newtheorem{prop}[thm]{Proposition}

\theoremstyle{definition}
\newtheorem{defn}[thm]{Definition}
\theoremstyle{remark}

\def\beq{\begin{eqnarray}}
\def\eeq{\end{eqnarray}}
\def\bsp{\begin{split}}
\def\esp{\end{split}}

\def\Tr{\mathrm{Tr}}

\newcommand{\Hxy}{{H(x,y)}}
\newcommand{\Hyx}{{H(y,x)}}

\newcommand{\Axy}{{A(x,y)}}
\newcommand{\Ayx}{{A(y,x)}}
\newcommand{\Lxy}{{L(x,y)}}
\newcommand{\Gx}{{G(x)}}
\newcommand{\Gy}{{G(y)}}

\newcommand{\Om}{{\Omega}}
\newcommand{\la}{{\lambda}}
\newcommand{\nn}{\nonumber}
\newcommand{\pd}{\partial}

\newcommand{\Csf}{{\sf C}}
\newcommand{\Rsf}{{\sf R}}
\newcommand{\Tsf}{{\sf T}}

\newcommand{\disc}[3]{{{}_{\sf #1}^{#2}\! D_{#3}}}
\newcommand{\DC}[1]{D_{#1}}
\newcommand{\DT}[1]{\disc{T}{5}{#1}}
\newcommand{\sign}{{\mathrm{sign}}}

\begin{document}
\title{\Large\textbf{Algebraic classification of five-dimensional spacetimes using scalar invariants }}
\author{{\large\textbf{A. A.~Coley$^{\heartsuit}$, 
                       S.~Hervik}$^{\text{\tiny\textleaf}}$\large\textbf{, 
                       M. N.~Durkee$^{\diamondsuit}$ and M.~Godazgar$^{\diamondsuit}$}} 
          \vspace{0.3cm} \\ 
          $^{\heartsuit}$Department of Mathematics and Statistics,\\ 
                         Dalhousie University, 
                         Halifax, Nova Scotia,\\ 
                         Canada B3H 3J5 
          \vspace{0.3cm}\\ 
          $^{\text{\tiny\textleaf}}$
          Faculty of Science and Technology,\\   
          University of Stavanger,\\  N-4036 Stavanger, Norway    
          \vspace{0.3cm} \\    
          $^{\diamondsuit}$DAMTP,\\ University of Cambridge,\\ 
          Centre for Mathematical Sciences, Wilberforce Road,\\ Cambridge, CB3 0WA, United Kingdom
          \vspace{0.3cm}\\
          \texttt{aac@mathstat.dal.ca, sigbjorn.hervik@uis.no,}\\
          \texttt{mark.durkee@cantab.net, M.M.Godazgar@damtp.cam.ac.uk} }   
\date{\today}   
\maketitle   
\pagestyle{fancy}   
\fancyhead{} 
\fancyhead[EC]{A.~Coley, S.~Hervik, M.~Durkee and M.~Godazgar}   
\fancyhead[EL,OR]{\thepage}   
\fancyhead[OC]{Classification of 5D spacetimes using scalar invariants}   
\fancyfoot{} 
   
\begin{abstract} 

There are a number of algebraic classifications of spacetimes in higher dimensions utilizing alignment theory, bivectors and discriminants. Previous work gave a set of necessary conditions in terms of discriminants for a spacetime to be of a particular algebraic type.  We demonstrate the discriminant approach by applying the techniques to the Sorkin-Gross-Perry soliton, the supersymmetric and doubly-spinning black rings and some other higher dimensional spacetimes.  We show that even in the case of some very complicated metrics it is possible to compute the relevant discriminants and extract useful information from them.

\end{abstract}  

\newpage

\section{Introduction}

Recently there has been considerable interest in the study of general relativity (GR) in higher dimensions and, in particular, in higher dimensional black holes \cite{HIGHER-D-REVIEW}.  The underlying motivation for this comes from supergravity, string theory and the gauge-gravity correspondence, but higher-dimensional GR has developed into a field of interest in its own right.

It has been shown that even at the classical level, gravity in higher dimensions exhibits much richer structure than in four-dimensions (4D).  One of most remarkable features of 4D GR is
the uniqueness of the Kerr black hole, but this fails in a striking way in higher dimensions.  
There now exist a number of different asymptotically flat, higher-dimensional vacuum black hole solutions \cite{HIGHER-D-REVIEW}, including Myers-Perry black holes \cite{MP}, black rings \cite{RBR,DSBR}, and various solutions with multiple horizons (e.g.\ \cite{Elvang:2007sat,Elvang:2007bi}).

The algebraic classification of spacetimes has played a crucial role in understanding black holes in 4D (for example, in the discovery of the Kerr metric, and the study of its linearized perturbations \cite{kramer}).  
Algebraic classification in 4D can be described in several different ways, using null vectors, 2-spinors, bivectors or scalar invariants.  There are strong links between these approaches; indeed, each of them can be used to give a different description of the 4D Petrov classification scheme.  Algebraic classification has been generalized to higher dimensions using each of these different methods \cite{desmet, class, BIVECTOR}, but it turns out that each approach leads to a distinct classification in higher dimensions \cite{spinorclass, BIVECTOR}.

For vacuum solutions, where the Ricci curvature tensor vanishes, algebraic classification reduces to the classification of the properties of the Weyl curvature tensor.  The most comprehensive and (to date) well-studied approach \cite{class,Alignment,AlignmentReview} classifies null frame components of the Weyl tensor according to their boost weights under local Lorentz boosts by identifying a choice of null frame such that components of high boost weight vanish.  More explicitly, one works in a particular frame 
\begin{equation}\label{vecbasis}
  \{\ell,n,m_i\}, \qquad\qquad i=2,3,\ldots,D-1 
\end{equation}
where $\ell$ and $n$ are linearly independent null vectors, transforming as
\begin{equation}
  \ell \mapsto \la \ell, \qquad n\mapsto \la^{-1} n, \qquad m_i \mapsto m_i
\end{equation}
We will refer to this as the \emph{alignment classification}.  This classification
divides spacetimes into 6 primary different types:  {\bf G}, {\bf I}, {\bf II}, {\bf III}, {\bf
N}, {\bf O} (for definitions and further details of these various algebraic types see,
for example, \cite{class,AlignmentReview,PraPraOrt07,higherghp}).

Examples of spacetimes that are algebraically special in this classification include
the type {\bf D} Myers-Perry black holes (i.e., they admit two
independent null vector fields with the type {\bf II} property).  Black ring spacetimes
are also algebraically special on the horizon \cite{brwands}; this will be discussed in
more detail below.  Applications of this classification scheme include the recent
construction of gauge invariant variables describing perturbations of any spacetime that
is of type {\bf II} or more special \cite{decoupling}.

Compared to four dimensions, the algebraic types defined by the higher-dimensional
alignment classification are rather broad, and it has proven more difficult to derive
general results.  Therefore, it is natural to ask whether there is a useful way of
refining these types further.
The higher-dimensional bivector classification \cite{BIVECTOR} achieves this
by analysing the bivector map
\footnote{In all of what follows, Greek indices $\alpha,\beta,\mu,\nu,\ldots$ are
$D$-dimensional spacetime indices.}
\begin{equation}\label{map:bivector}
 \Csf: X_{\mu\nu} \mapsto \tfrac{1}{2} C^{\phantom{\mu\nu}\rho\sigma}_{\mu\nu} X_{\rho\sigma},
\end{equation}
which maps the space of spacetime bivectors (2-forms) $X$ to itself.  
In four dimensions,
a decomposition into (anti-)self dual parts can be used to render the bivector
classification identical to the standard Petrov algebraic classification\cite{kramer}.  
However, in higher dimensions this is not possible.

In \cite{BIVECTOR}, two of the present authors analysed the bivector map in higher dimensions, and
showed how its eigenvalue structure is related to the alignment classification
\cite{class}.
In particular, they expressed the operator explictly in a manner consistent with its boost
weight decomposition, and then studied its eigenbivector and eigenvalue structure
in order  to refine the alignment classification 
(i.e., to divide several of the
algebraic types into subtypes).  The hope of such a refinement is that it might be possible
to prove
more powerful results in one or more of these more restricted types.\footnote{ Note
that the alignment types {\bf G} and {\bf I} are both of equal generality with respect
to their possible (eigenbivector/eigenvalue) roots structure. It has been
suggested for other reasons that the distinction between these types may not be
significant in certain applications \cite{mahdi}.}

From a computational point of view, a disadvantage of these two classification schemes is
that one must solve a complicated set of equations to find the `preferred' null frame (of
vectors or bivectors) in which to do calculations.  It would be very useful if there was
a more constructive way of accessing the invariant classification information.  In 4D,
such an approach is given by the use of scalar curvature invariants (e.g., see
\cite{kramer}).  In \cite{DISCRIM}, similar algebraic techniques were developed in higher
dimensions using `discriminants'.

As an example, consider the bivector operator $\Csf$ defined in (\ref{map:bivector}).
For a spacetime of algebraically special type, the operator $\Csf$ will have a restricted
eigenvector structure \cite{OP, BIVECTOR}.  That is, we can obtain necessary conditions
for a particular algebraic type, in terms of the eigenvalue structure of the matrix.

Studying eigenvalues of the bivector operator alone can only give so much information.
However, as we will discuss in Section \ref{sec:disc}, there are many other operators
that can be considered, constructed from various contractions of the Riemann tensor and
its derivatives.  It was shown in \cite{DISCRIM} that it is possible to derive necessary
conditions for a spacetime to be of various algebraic types in terms of these other
operators.

Computing eigenvalues of large matrices explicitly is difficult, but in fact there is no
need to do this in order to obtain the information that we require.  In \cite{DISCRIM},
two of the present authors showed how one can use an explicit algorithm to completely
determine the eigenvalue structure of the curvature operator, up to degeneracies, in
terms of a set of discriminants described in Section \ref{sec:disc} and in the Appendix.
Since the characteristic equation has coefficients which can be expressed in terms of the
scalar polynomial curvature invariants of the curvature tensor,~\footnote{ A {\em scalar
polynomial curvature invariant of order $k$} is a scalar obtained by contraction from a
polynomial in the Riemann tensor and its covariant derivatives up to the order $k$.}
these conditions (discriminants) can be expressed in terms of these polynomial curvature
invariants.

In particular, these techniques can be used to study the necessary conditions in arbitrary dimensions
for the Weyl curvature operator (and hence the higher dimensional Weyl tensor) to be of algebraic type
{\bf II} or {\bf D}, and create syzygies~\footnote{A \emph{syzygy} is an algebraic relationship between
scalar polynomial curvature invariants.}  which are necessary for the special algebraic type to be
fulfilled.  We are consequently able to determine the necessary conditions in terms of simple scalar
polynomial curvature invariants in order for the higher dimensional Weyl tensors to be of type {\bf II}
or {\bf D}.

Let ${\cal I}$ be the set of all scalar polynomial curvature invariants.  An important
question is whether the ${\cal I}$ is unique for the spacetime under consideration.  In
particular, does there exists a continuous deformation, $g_{\tau}$, of the metric $g$ so
that $g_{\tau}$ gives the same invariants as $g$?  If such a deformation exists we will
say that the metric is ${\cal I}$-degenerate (otherwise, it will
be said to be ${\cal I}$-non-degenerate).  In
4D this question was discussed in \cite{inv}, and it was found that the only spacetimes that 
are not  ${\cal I}$-non-degenerate are the (very special) degenerate Kundt spacetimes.  In particular, this implies that
black hole spacetimes are ${\cal I}$-non-degenerate and that the set of scalar invariants can be used to
characterise such spacetimes.  The same is believed to be the case in higher dimensions
and it is ${\cal I}$-non-degenerate spacetimes that will be the focus of study here.

The purpose of the current paper is twofold.  First, we demonstrate that the
discriminant approach is reasonably practical in that even in the case of some very
complicated metrics it is possible to compute the relevant discriminants and extract
useful information from them, at least with the use of computer algebra.  Unfortunately,
for some of the metrics that we discuss we will find that doing computations for generic
values of various parameters is too difficult, and hence we will be forced to study
particular representative examples within families of spacetimes (for example, in the case
of the Pomeransky-Sen'kov doubly spinning black ring \cite{DSBR}).

Second, we look to better understand the links between these discriminants and the
alignment type of spacetimes.  Previous work \cite{DISCRIM} gave a set of necessary (but
not sufficient) conditions on the discriminants for a spacetime to be type {\bf II} (or
more special).  Here we illustrate these conditions in a number of examples, and also
prove a partial result (Proposition \ref{prop:isotropy}) that demonstrates that stronger
conditions hold in the case of spacetimes with considerable isotropy.
Given these additional conditions in the case of spacetimes with isometries (at least for
some set of points in the spacetime), we are able to see that in all cases discussed in
this paper the necessary conditions relevant to the isometry group of the spacetime also
turn out to be sufficient.  This naturally leads one to speculate that there might exist
a more complete result linking scalar invariants and alignment classification more
closely.

A useful application of these techniques to the alignment classification of spacetimes is
also given.  It is known \cite{brwands} that the singly-spinning black ring \cite{RBR} is of 
type {\bf II} on the horizon, but only of type {\bf I} or {\bf G} outside.  The relevant
discriminants were computed in \cite{DISCRIM} and found to be consistent with this
result.  Therefore, it might be expected that the doubly-spinning black ring would also
have this property.  Here, we verify this by showing that the relevant discriminants are
non-vanishing outside the horizon,~{\footnote{To be precise, 
the discriminants  are computed at a number of particular parameter values and for fixed values of one of the coordinates;
see Section 5.}}
and hence that the doubly-spinning black ring must be of
type {\bf G} or {\bf I} in the exterior region (it is easy to check, for example by
explicit calculation in Gaussian null coordinates \cite{Moncrief:2008}, that all black
holes are of type {\bf II} or more special on the horizon).

\newpage
\section{Review of the Discriminant Approach} \label{sec:disc}

In this section we review the scalar invariant approach to the algebraic classification of
spacetimes introduced in \cite{DISCRIM}, which links scalar invariants and bivector
operators \cite{BIVECTOR} with the alignment classification \cite{class}.  The approach
described in \cite{DISCRIM} was based on a set of algorithms for determining the
structure of roots of a polynomial, derived in \cite{Yang:1996,Yang:1997,Liu} (more
details of these results are given in the Appendix).

We will develop this work further in the case of spacetimes with isometries, proving a
new result that extends the usefulness of scalar invariant methods in this
case.

\subsection{Curvature operators}
Given a curvature tensor, an operator describing an automorphism of
some finite-dimensional vector space can be defined.  The most well-known example of such a map is the
bivector map (\ref{map:bivector}) obtained from the Weyl tensor, mapping the $n =
D(D-1)/2$-dimensional space of 2-forms to itself.

If the Weyl tensor is of a particular algebraic type in the alignment classification, then the associated operator $\Csf$ will have a restricted eigenvector structure \cite{BIVECTOR}.  One can extend this technique to include other operators, constructed either from the Weyl tensor, or from other curvature operators $\Rsf \in {\cal R}$, where ${\cal R}$ denotes the set of all curvature operators constructed from the curvature tensors and their polynomial invariants.  For example, a second operator that can be constructed from the Weyl tensor is given by
\begin{equation}
 {\tilde {\Csf}} : Y_{\mu\nu} \mapsto \tfrac{1}{2}
         C^{\phantom{\mu}\rho\phantom{\nu}\sigma}_{\mu\phantom{\rho}\nu} Y_{\rho\sigma}.
\end{equation}
acting on the $D^2$-dimensional vector space of 2-tensors $Y$, while the tensor
\begin{equation}
  T^{\alpha}_{~\beta} \equiv C^{\alpha\mu\nu\rho}C_{\beta\mu\nu\rho}
\end{equation}
can be used to construct an operator 
\begin{equation}  \label{eqn:Tdef}
  \Tsf: X^\alpha \mapsto T^{\alpha}_{~\beta} X^\beta  
\end{equation}
acting on the $D$-dimensional tangent space of the spacetime (in fact, it is often more useful to consider the operator constructed from the tracefree part of $T$).  

\subsection{Discriminant analysis}
For a given curvature operator, ${\sf R}$, we can consider the eigenvalues of this operator to obtain necessary conditions on various alignment types of the spacetime.  In particular, requiring the algebraic type to be {\bf II} or {\bf D}  will impose restrictions of the eigenvalues on the operator \cite{DISCRIM}.  
Consider the characteristic equation
\begin{equation} 
  f(\la) \equiv \det (\Rsf - \lambda {\sf 1} ) = 0.  
\end{equation} 
This equation is a polynomial equation in $\lambda$ and the eigenvalues are the roots of this equation.  The coefficients can be expressed in terms of the invariants of $\Rsf$.  Therefore, we can give conditions on the eigenvalue structure expressed manifestly in terms of the polynomial curvature invariants (syzygies) of ${\sf R}$.  

The precise details of how this can be done are given in the Appendix.  In the remainder of this section we briefly summarize the minimal information required to give a basic understanding of the language and notation used in the remainder of the paper.

Given a curvature operator, $\Rsf$, acting on an $n$-dimensional vector space, we can construct a unique \emph{discriminant sequence} for the characteristic polynomial $f$, consisting of $n$ scalar functions on spacetime, that we will denote by
\begin{equation} \label{def:discseq}
  \left[\disc{R}{n}{1},\disc{R}{n}{2},\ldots,\disc{R}{n}{n}\right].
\end{equation}        
We will make particularly regular use of the discriminant sequence for $\Csf$ in 
what follows, and for convenience we will abbreviate these discriminants by 
writing $\DC{i} \equiv \disc{C}{n}{i}$ throughout this paper.
\footnote{Ref.\ \cite{DISCRIM} used $\mathcal{CHP}$ to represent $\DC{10}$, 
$\DC{9}$, $\DC{8}$; to avoid introducing yet more notation we do not do this here.} 

Given a discriminant sequence (\ref{discseq}), its \emph{sign list} $S$ is defined to be 
\begin{equation}
  S = \left[\sign(\disc{R}{n}{1}),\sign(\disc{R}{n}{2}),\ldots,\sign(\disc{R}{n}{n}) \right]
\end{equation}
(where $\sign(x) \equiv 1,0,-1$, respectively, as $x>0, x=0, x<0$).

Given any sign list $S=[s_1,\ldots,s_n]$, we construct a \emph{revised sign list} 
$\bar{S}$ as follows.  First, we look for ``internal zeros'' of $S$ (that 
is, for any subsequences of the form $[s_i,0,0,\ldots,0,s_j]$, 
where $s_i\neq 0$ and $s_j\neq 0$).  If there are none, then we take $\bar{S}=S$.  
Otherwise, we replace any such subsequence with 
\begin{equation}
  [s_i,-s_i,-s_i,s_i,s_i,-s_i,-s_i,s_i,s_i\dots,s_j] 
\end{equation} 
in $\bar{S}$.  The revised sign list will therefore contain no ``internal'' zeros, but may have zeros at the end.

This is useful because of the following result:
\begin{lem}[\cite{Yang:1996,Yang:1997,DISCRIM}]\label{lem:roots}
Consider the revised sign list $\bar{S}$ of the discriminant sequence $\left[\disc{R}{n}{1},\disc{R}{n}{2},\ldots,\disc{R}{n}{n}\right]$.  Let $K$ denote the number of sign changes and $L$ denote the number of non-zero members of $\bar{S}$.
Then, the number of distinct \emph{pairs of complex conjugate roots} is $K$, and the number of distinct \emph{real roots} is $L-2K$. 
\end{lem}

Hence, by applying the algorithm given in the Appendix to compute $\bar{S}$ (which can
conveniently be done in an automated way using a computer algebra
system such as, for example, Maple), we
can learn a significant amount about the eigenvalue structure of $\Rsf$.

\subsection{Necessary conditions for Weyl type {\bf II}/{\bf D}} 

As discussed in the introduction, for the Weyl tensor to be type {\bf II} (or more
special) the eigenvalues of the corresponding bivector operator $\Csf$ need to be of a
special form.

In fact, since the invariants of a type {\bf II} spacetime are 
the same as for type {\bf D}, we will assume for simplicity that the spacetime 
is type {\bf D} for the purposes of the present discussion.  In \cite{BIVECTOR}, 
it was shown that in this case we can choose a basis
\begin{equation} \label{bivectorbasis}
  \ell \wedge m_i, \qquad
  \ell \wedge n, \qquad
  m_i \wedge m_j, \qquad
  n \wedge m_j
\end{equation}
adapted to the vector basis (\ref{vecbasis}).  Hence, the space of bivectors can be 
split into three vector subspaces of boost weights $+1$, $0$ and $-1$, of dimensions 
$(D-2)$, $(n-D+2)$ and $(D-2)$, respectively.  With respect to this basis, the bivector operator takes the form of an $n\times n$ matrix ($n = D(D-1)/2$)
\begin{equation}
  \Csf = \mathrm{blockdiag}(M,\Psi,M^t),
\end{equation}
where $M$ is a $(D-2)\times(D-2)$ matrix and $\Psi$ is an $(n-D+2)\times(n-D+2)$ 
matrix \cite{DISCRIM}.  Since the eigenvalues of $M$ and $M^t$ are the same, 
it must be the case that:~\footnote{We have discussed the type {\bf D} case here. 
In the type {\bf II} case the block-diagonal form above reduces to a `block lower-triagular' 
form for which a similar argument applies.  
The case of a spacetime that is more special than this (i.e.,  primary type {\bf III}, {\bf N} or {\bf O}) is far simpler; this occurs if and only if the bivector operator is nilpotent.}
\begin{lem}[\cite{BIVECTOR,DISCRIM}] 
   The Weyl bivector operator for a spacetime that is of type 
{\bf II} or more special has at least $(D-2)$ eigenvalues of multiplicity (at least) 2.
\end{lem}
                              
Consider now the particular case of five dimensions.  Here, the bivector operator acts on
a vector space that is $n=10$-dimensional and if the spacetime is type {\bf II} then the
bivector operator has 3 eigenvalues of (at least) multiplicity 2.  Hence the eigenvalue
(Segre) type of the matrix is $\{(11)(11)(11)1111\}$ (or more special, e.g.\
$\{(1111)(11)1111\}$). From the point of view of the discrimant analysis, it was 
shown in \cite{DISCRIM} that eigenvalue types consistent with at least 3 matching 
pairs of eigenvalues can only occur if the last three discriminants in the sequence 
(\ref{def:discseq}) vanish (see the Appendix for further details).  Hence, we arrive at the most important result for this paper:
\begin{prop}[\cite{DISCRIM}]\label{prop:main}
  If a five-dimensional spacetime is of alignment type {\bf II} or more special, then
  \begin{equation}
    \DC{8} = 0, \qquad
    \DC{9} = 0, \qquad 
    \DC{10} = 0.
  \end{equation}  
\end{prop}
These three equations are syzygies of order 90, 72 and 56, respectively.  In principle,
these can be computed in general, and written out in full in terms of the Weyl tensor.
In reality, this is  not practical, and it is far more useful to apply the algorithm
used to construct the discriminants in the case of particular metrics.

In this paper we shall also consider the operator $\Tsf$ defined in (\ref{eqn:Tdef}) above.  Following a similar approach to the bivector case, it can be shown that: 
\begin{prop}[\cite{DISCRIM}]
  For a five dimensional spacetime of algebraic type {\bf II} or more special, the discrminants associated with the operator $\Tsf$ satisfy:
  \begin{equation}
    \DT{5}=0, \qquad
    \DT{4} \geq 0, \qquad
    \DT{3} \geq 0, \qquad 
    \DT{2} \geq 0.
  \end{equation} 
\end{prop}
Note that $\DT{5}=0$ is a 40th order syzygy in the Weyl tensor (a 20th order syzygy in the square of the Weyl tensor), and hence is likely to be easier to calculate explicitly than the syzygies resulting from the bivector operator.  

\subsection{Spacetimes with isometries}
Note that Proposition \ref{prop:main} gives necessary conditions for a spacetime to be
type {\bf II} or more special. These conditions are not sufficient.  
Indeed, the necessary conditions of Proposition 2.3 
may be fulfilled in spite of the fact that a spacetime is of type {\bf G} or {\bf I}.
Many examples of the
non-sufficiency of this condition can be found by considering spacetimes with large
amounts of symmetry; we will see several examples of this in the remainder of the paper.
However, in this section, we will show how for spacetimes (or sets of points in
spacetimes) admitting an $SO(d-2)$ isometry, there is a stronger necessary condition on
the discriminants which may be of more use.

More explicitly, it was shown in \cite{BIVECTOR} that a 5D spacetime with a Weyl tensor
that has an $SO(2)$ isotropy with spacelike orbits must automatically have 3 pairs of
matching eigenvalues.  Intuitively, this is because one can choose a basis in which the
isometry makes all quantities invariant under $SO(2)$ transformations mixing $m_2$ and
$m_3$, and hence both the $3\times 3$ matrix M, and the $4\times 4$ matrix $\Psi$ must
have two of their eigenvalues the same.  Matching this with the result above, we see that
one eigenvalue must be repeated at 4 times (since $M$ has a repeated eigenvalue, and
$M^t$ has the same repeated eigenvalue), and hence the eigenvalue type of the bivector
matrix must be $\{(1111)(11)(11)11\}$ or more special.  This is now a basis independent
condition.

This condition  can be interpreted in terms of the discriminants $\DC{i}$; thus 
\begin{prop} \label{prop:isotropy}
  If a 5D spacetime possesses a spatial $SO(2)$ isotropy, and is of alignment type {\bf II} or more special, then 
  \begin{equation}
     \DC{7} = \DC{8} = \DC{9} = \DC{10} = 0. 
  \end{equation} 
\end{prop}
Further details of this are given in the Appendix (of course the second half of the equalities follow trivially from Proposition \ref{prop:main}; but the important point is that here they also hold as a result of the isometry itself, and hence we always need some additional conditions).

Note that this is a pointwise result, and hence it also holds on lines of enhanced symmetry within a spacetime.  The most important example of this is on an axis of rotation in an axisymmetric spacetime; we shall see several examples of this in the sections that follow.

\subsection{A caveat: Discrete symmetries}
Leaving aside the links to alignment classification for the moment, the bivector methods \cite{BIVECTOR} also provide an independent algebraic classification in their own right; with algebraic types labelled by the eigenvalue (Segre) types of the bivector operator.

Using the invariants only, we can only determine the \emph{eigenvalue type} of the operator.  For example, 
we cannot distinguish between the eigenvalue types  $\{(1,1)11 ... 1\}$ and $\{1,1 ... 1(11)\}$.  This is a result of the fact that the spacetime is Lorentzian and that time and space directions have a different interpretation in spacetime. The discriminants cannot \emph{a priori} distinguish the time and space directions. This is a caveat that is distinct to the Lorentzian nature of spacetime and  is not present in analyses of Riemannian (i.e., signature (+++..+) ) spaces. 
Note also that this degeneracy is a \emph{discrete} degeneracy, unlike the notion of $\mathcal{I}$-degenerate metrics \cite{inv} which requires a \emph{continuous} deformation.

\section{The Sorkin-Gross-Perry soliton} 

In order to illustrate some of these caveats, we now consider the example of the Sorkin-Gross-Perry (SGP) soliton, a class of 5-dimensional, vacuum solutions with a large amount of symmetry \cite{soliton}.  

The 5D SGP metric is given by:
\begin{eqnarray}
  ds^2 &=& - \left( \frac{1-m/R}{1+m/R} \right)^{2/\alpha} dt^2 
         + \left( \frac{1-m/R}{1+m/R} \right)^{2\beta /\alpha} dz^2 \nonumber \\
       & & +  \left( 1+ \frac{m}{R} \right)^4 
            \left( \frac{1-m/R}{1+m/R} \right)^{2(\alpha-\beta-1)/\alpha} 
            \left( dR^2 + R^2 d\Omega^2 \right),\label{SGP}
\end{eqnarray}
where $\alpha \equiv \sqrt{\beta^2 + \beta +1}$ (for a vacuum soliton solution) and we assume that $m \neq 0$ (so the spacetime is not flat; in the calculations below we set $m=1$, which simply re-scales $R$), and the coordinates range over $-\infty < t,z < \infty$ and $m<R<\infty$.

Generically, the solution does not contain a black hole; it closes off smoothly at $R=m$.   However, there are a number of special cases:
\begin{enumerate}
  \item[(i)] If $\beta=0$, it is the Schwarzschild black string, which is of type {\bf D} everywhere. 
  \item[(ii)] If $\beta=1$, it is a boost invariant singular spacetime of type {\bf  G}.
  \item[(iii)] The limit $\beta \rightarrow \infty$ corresponds to the static Kaluza-Klein bubble; that is, the product of a flat time direction with the Euclidean 4D Schwarzschild solution, which is of type {\bf{G}} everywhere.  
\end{enumerate}

In other cases, these solutions are of Weyl alignment type {\bf G} or {\bf I}
\cite{5Dclass,mahdi}.  In fact, it was noted in \cite{mahdi} that these provide examples
of connected analytic spacetimes that are type {\bf{G}} in some open region and type
{\bf{I}} in some other open subset, determined by the sign of a particular inequality.
For example, if we choose parameters $m>0$ and $\beta = 1/2$ in the above metric (which
exhibits behaviour typical for the case $0< \beta <1$ and $m>0$), then it is of type
{\bf{G}} for $R \sim m$ but of type {\bf{I}} for $R \gg m$.  Conversely, for (finite)
$\beta>1$ and $m>0$, it is of type {\bf I} near $R=m$ and of type {\bf{G}} for $R \gg
m$).  \footnote{By solving a quartic equation it was claimed in \cite{5Dclass} that a
null coframe exists in which the positive $+2$ (and negative $-2$) boost weight Weyl
components vanish, and hence that the SGP metric is of type {\bf{I}} ($I_i$) everywhere.
However, it was not noticed that the solutions for the roots of this quartic are not
always real, and in some regions in which the roots are complex the algebraic type is in
fact {\bf{G}} (i.e., the preferred null vector is actually complex).  Note that within
the bivector formalism \cite{BIVECTOR} these solutions are of equivalent algebraic
specializations in the different regions.  This is consistent with the results of
\cite{mahdi} discussed above and corrects the omission in \cite{5Dclass}.}

This spacetime is static: $\pd/\pd t$ is a hypersurface orthogonal timelike Killing vector field.  It was shown in \cite{PraPraOrt07} that all static spacetimes (and a large class of stationary spacetimes) in dimensions $D > 4$ are necessarily of alignment type {\bf G}, {\bf I}$_i$, {\bf D} or {\bf O}.  This implies that this spacetime cannot be of type {\bf II}, {\bf III} or {\bf N}.  Essentially, the reason for this is that static spacetimes allow for the discrete transformation $t\mapsto -t$ which interchanges the two null directions. 

\subsection{Discriminant Analysis} 

We now look to compute the discriminant sequence for the bivector operator $\Csf$ for this metric.
For all of the SGP spacetimes (with $R \geq m$), we find that $\DC{1}=10$ (which merely sets a normalization) but also that 
\begin{equation}
  \DC{8} = \DC{9} = \DC{10} = 0.
\end{equation}
Hence, the necessary condition from Proposition \ref{prop:main} for the spacetime to be of alignment type {\bf II} or more special is always met.  However, this spacetime has sufficient symmetry that the necessary conditions of interest here are the stronger ones of Proposition \ref{prop:isotropy}. 
In the general case, all other $\DC{i}$ are non-zero, and these stronger conditions are not satisfied.

To see this in more detail, consider a typical case $\beta = 1/2$ (and, without loss of generality, set $m=1$).  For convenience, define $Q(R) \equiv -7R^2 -7 +6 \sqrt{7}R$, which only has one real root with $R \geq 1$ (namely $R_0 = (3 \sqrt{7} + \sqrt{14})/7$).  We then find that, for $R$ close to $R_0$, we have that
\begin{equation}
  \DC{2} > 0, \qquad 
  \DC{3} >0 ,
\end{equation}
and
\begin{equation}
  \DC{4} = (+ve)Q^2, \quad
  \DC{5} = (+ve)Q^4, \quad
  \DC{6} = (+ve)Q^6, \quad 
  \DC{7} = (+ve)Q^{10}.
\end{equation}

Hence, for $R\neq R_0$ (i.e., $Q\neq 0$) the sign list (and revised sign list) of this
operator is $[1,1,1,1,1,1,1,0,0,0]$.  This contains no changes of sign, and hence Lemma
\ref{lem:roots} tells us that the operator has no complex eigenvalues, and 7 distinct
real eigenvalues.  Therefore, the eigenvalue type of the 5D Weyl tensor is
$\{(11)(11)(11)1111\}$, corresponding to type {\bf  I/G} (consistent with the results of 
\cite{mahdi} and recalling that types {\bf{G}} and {\bf{I}} are of equivalent algebraic specialization
within the bivector formalism \cite{BIVECTOR}). Note that even with 
$\DC{8} = \DC{9} = \DC{10}= 0$, this eigenvalue structure 
implies type {\bf{G}} or {\bf{I}}
because the  SGP spacetime has a spatial isotropy (as discussed above and in \cite{OP}).

Note that on $R=R_0$, $\DC{4} = \DC{5} = \DC{6} = \DC{7} = 0$, and so on this line the
spacetime could be of type {\bf D}.  This is indeed the case \cite{mahdi}.

Let us next consider the special cases $(i)-(iii)$. 
\begin{enumerate}
  \item[(i)] In the case of the black string ($\beta =0$), we have that
             \begin{equation}
               \DC{1} = 10,
               \DC{2} > 0,
               \DC{3} > 0,
             \end{equation}
             and
             \begin{equation}
               \DC{4} = \DC{5} = \DC{6} = \DC{7} = \DC{8} = \DC{9} = \DC{10} = 0.
             \end{equation}   
             This satisfies the necessary condition for a spacetime with an $SO(2)$ 
             isometry to be algebraically special and, in fact, the black string spacetime is of type {\bf{D}} everywhere.
  
  \item[ii)] For $\beta = 1$ we have that $\DC{i} > 0$, for $i=1-5$, and $\DC{i} = 0$, for $i=6-10$. As noted above, this case is of type {\bf  G}.
  
  \item[(iii)] The static KK bubble spacetime $\beta \rightarrow  \infty$ also has 
             \begin{equation}
               \DC{1} = 10, \DC{2} >0 , \DC{3} >0, 
             \end{equation}
             and
             \begin{equation}
               \DC{4} = \DC{5} = \DC{6} = \DC{7} = \DC{8} = \DC{9} = \DC{10} = 0,
             \end{equation}  
             and is of type {\bf G} everywhere.
\end{enumerate} 
For the black string and the static KK bubble, $\DC{1}$, $\DC{2}$, $\DC{3}$ are positive, while all other $\DC{i}=0$, but the black string is of type {\bf D} and the static KK bubble is of type {\bf G}.  However, there is a discrete symmetry (interchanging $r$ and $t$) relating these two 5D spacetimes (which are related by a complex continuation) and hence their invariants are identical but their Weyl types are different (this is similar to the discrete, conformally flat, example given in \cite{inv}).

\section{Supersymmetric Black Ring}

The results of this paper do not only apply to vacuum spacetimes.  To emphasize this, we now move on to consider an interesting class of non-vacuum metric, corresponding to supersymmetric, rotating black holes in five dimensions.  The simplest example of such a spacetime is the extremal, charged, rotating BMPV black
hole \cite{BMPV} in 5D minimal supergravity, which has a horizon of spherical topology.  It is already known that the BMPV metric is generally of Weyl type {\bf I}$_i$ \cite{5Dclass}.  

Later, a supersymmetric black ring solution of the same supergravity theory was presented in \cite{SBR}, with a black hole horizon of topology $S^1 \times S^2$.  In fact, this solution can be seen as a generalization of the BMPV metric, which can be obtained as a particular limit of the supersymmetric black ring family.  

\subsection{Metric and properties}

Since the BMPV metric is a particular case of the supersymmetric black ring (having a horizon topology $S^{1}\times S^{2}$), one would expect that the supersymmetric black ring is not of Weyl type {\bf II} or more special, but this has not been proved.  This question is something that we can attack with relative ease using these discriminant methods.

The line element of the supersymmetric black ring is
\begin{equation}
  ds^2= -f(x,y)^2 (dt+\omega)^2 + f(x,y)^{-1} ds^{2}(\mathbf{R}^{4}),
\end{equation}
where
\begin{eqnarray}
  f(x,y)^{-1}  &\equiv & 1+\tfrac{Q-q^2}{2R^2}(x-y)-\tfrac{q^2}{4R^2}(x^2-y^2),\\ 
  \omega       &=      & \omega_{\phi}d\phi+\omega_{\psi}d\psi , \label{deff} \\
  \omega_{\phi}&=      & -\tfrac{q}{8R^2}(1-x^2)[3Q-q^{2}(3+x+y)], \\
  \omega_{\psi}&=      & \tfrac{3}{2}q(1+y)+\tfrac{q}{8R^2}(1-y^2)[3Q-q^{2}(3+x+y)],
\end{eqnarray}
and the four dimensional flat space metric is written as
\begin{equation}
  ds^2(\mathbf{R}^{4}) = \frac{R^2}{(x-y)^2}\left[\frac{dy^2}{y^2-1}+(y^2-1)d\psi^2
                                                  +\frac{dx^2}{1-x^2}+(1-x^2)d\phi^2\right].
\end{equation}
Admissible coordinates values are $-1\leq x \leq 1$, $-\infty < y \leq -1$ and $\phi$,$\psi$ are $2\pi$-periodic; and the parameters $q$ and $Q$ satisfy $q>0$ and $Q \geq q^2$ (which implies that $f(x,y)>0$).

The black ring horizon lies at $y=-\infty$ in these coordinates.  Of course a different choice of coordinates can be made that includes the black hole horizon; and typically one must do this when trying to do any computation of a quantity on the horizon.  A great advantage of scalar invariant approaches is that this is not necessary; one can compute a scalar invariant everywhere outside the horizon and then extend it onto the horizon itself by continuity.  Hence, we are able to work in this relatively simple coordinate system throughout.

\subsection{Discriminant Analysis}

Unfortunately, computing the discriminants $\DC{i}$ explicitly for arbitrary values of
the parameters has proven to be difficult; hence we are forced to pick particular values of
the parameters (i.e., particular spacetimes in the family) to analyse.  The parameter $R$
merely fixes the length-scale, and hence we can set $R=1$ without loss of generality.  We
have then computed the discriminants for various values of the parameters $q$ and $Q$;
for the purposes of the present discussion, we consider the case
$q=1/2$, $Q=9/4$. {{\footnote{ We also  computed the discriminants   in the cases
$q=1/4$, $Q=1$; $q=1/2$, $Q=1/2$; $q=1$, $Q=2$.}}

Let us first consider the Weyl operator on the horizon $y=-\infty$.  For the Weyl bivector operator $\Csf$, the discriminants can be evaluated, and give
\begin{equation}
  \DC{1} = 10, \ 
  \DC{2} = 180, \
  \DC{3} = 4608; \ \
  \DC{4} = \DC{5} = \DC{6} = \DC{7} = \DC{8} = \DC{9} = \DC{10} = 0.
\end{equation}
Hence, this is consistent with the black ring being type {\bf II} or {\bf D} on the horizon, as we know it must be.
Explicitly, the eigenvalue type of the Weyl bivector operator is of the form $\{1(111)(111111)\}$ on the horizon, while 
the operator $\Tsf$ is of eigenvalue type $\{(11)(111)\}$ there.

Outside the horizon, the computations are rather more complicated.  Computing the operator $\Tsf$ indicates that the discriminant $\DT{5}$ takes the form:
\begin{equation}
  \DT{5} = F(x,y) \frac{(y-1)^2 (y+1)^2 (x-1)^2 (x+1)^2 f(x,y)^{154}}{(x-y)^{70}},
\end{equation}
where $F(x,y)$ is some (known) polynomial in $x$ and $y$ and $f$ is the metric function defined in (\ref{deff}).  Recall that wherever $\DT{5}$ is non-zero, the Weyl tensor must be of type {\bf I} or {\bf G}.

Checking that $F(x,y)$ is non-zero is relatively difficult.  We have verified that this is true for the following cases (among others): 
\begin{itemize}
  \item $x=0$, $y$ general. 
  \item $x=\pm 1/2$, $y$ general. 
  \item $y=-2$, $x$ general.
\end{itemize}
This shows that the metric is of type {\bf I/G} except possibly for a few special values of
$(x,y)$. 
For the SUSY black ring, some of the special cases are as follows:  $x=y$, which
represents a curvature singularity, occurs only at x=y=-1 
(for the coordinate ranges used here) and corresponds to asymptotic infinity.  $x=\pm 1$ 
is the plane of the ring
(the axis of $\phi$ rotation) and $y=-1$ is the axis of $\psi$-rotation.

\section{Doubly-spinning black ring}

Pomeransky \& Sen'kov \cite{DSBR} constructed a family of exact solution to the vacuum Einstein equations in 5D, corresponding to a black ring with both angular momenta non-vanishing.  The algebraic type of this metric is currently unknown.  Since this family contains the Emparan-Reall black ring \cite{RBR} as a special case, it seems unlikely that other doubly-spinning black rings are more algebraically special than this solution.  To prove this directly, one must demonstrate the non-existence of solutions to the appropriate alignment equations; the difficulty in doing this lies in the complexity of the metric.  

Therefore, as a demonstration of the utility of the discriminant techniques, let us do this by another method, and using the discriminant techniques discussed above prove that the doubly spinning black ring is not type {\bf II} or more special, except on the horizon.  Unfortunately, we are still not able to do this explicitly for general parameters $R$, $\la$, $\nu$ (defined below), as the resulting equations are too complicated even with computer algebra, but we are able to do it for a representative sample of different parameters covering all known cases of qualitatively different behaviour.

\subsection{Metric and parameters}
The metric can be written in the form \cite{ringgeo}:
\begin{multline} \label{eqn:metric}
  ds^2 = -\frac{\Hyx}{\Hxy}(dt+\Omega)^2  
        + \frac{R^2 \Hxy}{(x-y)^2 (1-\nu)^2} \left[\frac{dx^2}{\Gx}-\frac{dy^2}{\Gy} \right. \\
                    \left.   +\frac{\Ayx d\phi^2-2\Lxy d\phi d\psi- \Axy d\psi^2}{\Hxy \Hyx}
                                              \right],
\end{multline}
where the coordinates lie in the ranges $-\infty <t<\infty$, $0\leq \phi,\psi <2\pi$, $-1\leq x \leq 1$, $-\infty < y \leq -1$.  The metric is globally asymptotically flat, although this is not manifest in these coordinates, where asymptotic infinity corresponds to a point $(x,y)=(-1,-1)$. 

The one-form $\Omega$ characterizing the rotation is given by $\Om = \Om_\psi(x,y) d\psi + \Om_\phi(x,y) d\phi$, where
\begin{multline}
  \Om_\psi = -\frac{R\la\sqrt{2((1+\nu)^2-\la^2)}}{\Hyx} \frac{1+y}{1-\la+\nu}
                                               \big(1+\la-\nu+x^2 y \nu (1-\la-\nu) \\
                                                     + 2\nu x(1-y)\big)
\end{multline}
and
\begin{equation}
  \Om_\phi = - \frac{R\la\sqrt{2((1+\nu)^2-\la^2)}}{\Hyx} (1-x^2)y\sqrt{\nu},
\end{equation}
while the polynomial functions $G$, $H$, $A$, $L$ read
\begin{eqnarray}
  \Gx  &=&(1-x^2)(1+\la x + \nu x^2),\\
  \Hxy &=& 1+\la^2-\nu^2+2 \la \nu(1-x^2)y+2 x \la(1-y^2 \nu^2) \nn\\
       &=& +x^2 y^2 \nu (1-\la^2-\nu^2),\\
  \Lxy &=& \la \sqrt{\nu} (x-y)(1-x^2)(1-y^2) \big[1+\la^2-\nu^2+2(x+y)\la \nu \nn\\
       & & \quad\quad\quad\quad\quad\quad\quad\quad\quad\quad\quad\quad\quad - xy\nu(1-\la^2-\nu^2)\big],\\
  \Axy &=& \Gx(1-y^2)\left[((1-\nu)^2-\la^2)(1+\nu)+y\la(1-\la^2+2\nu-3\nu^2)\right]\nn \\
       & & + \Gy\big[ 2\la^2+x \la((1-\nu)^2+\la^2)+x^2((1-\nu)^2-\la^2)(1+\nu) \nn\\
       & & \quad \qquad + x^3 \la(1-\la^2-3\nu^2+2\nu^3)+x^4\nu(1-\nu)(1-\la^2-\nu^2)\big]. \hspace{6mm}
\end{eqnarray}

The (outer) horizon lies at $G(y)=0$ or
\begin{equation}
  y=y_h \equiv \frac{-\lambda+\sqrt{\lambda^2-4\nu}}{2 \nu}.
\end{equation}
In this paper, we will not consider the black hole interior.

The parameter $R>0$ sets the length-scale in the spacetime, while the dimensionless parameters $\lambda$ and $\nu$ are restricted to
\begin{equation}\label{twospinrange}
  R > 0, \qquad 
  0\leq\nu<1\,,\qquad 
  2\sqrt{\nu}\leq\lambda<1+\nu.
\end{equation} 
When $\lambda=\nu=0$ the metric reduces to that of flat spacetime, written in ring-like coordinates.  The singly spinning limit is obtained by setting $\nu = 0$.\footnote{To obtain the precise form of the singly spinning black ring given in the review articles \cite{HIGHER-D-REVIEW, ER:2006}, one must identify $R^2= 2k^2 (1+\lambda)^2$ and rename $\lambda\to \nu$.}  The bound $\lambda\geq 2\sqrt{\nu}$ corresponds to a Kerr-like upper bound on the rotation of the black ring around the $S^2$.  When $\lambda= 2\sqrt{\nu}$, the horizon is degenerate and we obtain an extremal black ring.

It was shown in \cite{Elvang:2008erg,ringgeo} that the properties of the black ring (outer) ergosurface vary drastically with the values of $\la$ and $\nu$.  For $2\sqrt{\nu} \leq \la < 1-\nu$, the ergoregion has the same topology as the event horizon ($S^1\times S^2$), but when $1-\nu < \la < 1+\nu$ the ergosurface has topology $S^3\cup S^3$, i.e.\ it consists of the union of two disjoint spheres, one surrounding the entire horizon, and the other lying in the centre of the ring.  There is a critical case at $\la = 1-\nu$ where the ringlike ergosurface just `pinches off' on an $S^1$. 

There is a generalization of this family that allows for a `non-equilibrium' black ring, supported by a conical singularity.  In this case, the metric functions take an extremely complicated form in the general case in which the black ring is not in equilibrium \cite{Morisawa}.  The extremal MP solution can be recovered as a limit of the extremal, non-eqilibrium solutions in this family, with $\nu\to 1$, $\lambda\to 2$.

\subsection{Discriminant Analysis}
We now look to evaluate the discriminants $\DC{i}$ and $\DT{i}$ for this metric.  Unfortunately, computing the determinants for general values of the parameters $\la$, $\nu$, and the coordinates $x$ and $y$, turns out to be too complicated, even with the use of Maple.

We will therefore compute them for particular solutions within the black ring family.  There are (at least) five distinct parts of parameter space $(\la,\nu)$ that give some kind of known, qualitatively different behaviour for the solution.  To try to ensure that we do not miss any important behaviour, we include an analysis of each of these cases:
\begin{itemize}
  \item {\bf Case 1. $0<\la<1$, $\nu=0$:} Singly spinning, non-extremal, $S^1\times S^2$ ergosurface (e.g.\ $\la=7/9$, $\nu=0$).
  \item {\bf Case 2. $2\sqrt{\nu}<\la<1-\nu$, $\nu>0$:} Doubly spinning, non-extremal, $S^1\times S^2$ ergosurface (e.g.\ $\la=7/9$, $\nu=1/9$).
  \item {\bf Case 3. $\la = 1-\nu$, $\nu>0$:} Doubly spinning, non-extremal, `pinched' ergosurface (e.g.\ $\la=8/9$, $\nu=1/9$).
  \item {\bf Case 4. $1-\nu<\la<1+\nu$, $\nu>0$:} Doubly spinning, non-extremal, $S^3\cup S^3$ ergosurface (e.g.\ $\la=1$, $\nu=1/9$).
  \item {\bf Case 5. $\la=2\sqrt{\nu}$, $\nu>0$:} Doubly spinning, extremal (e.g.\ $\la=2/3$, $\nu=1/9$).
\end{itemize}

First, we consider the discriminants $\DC{i}$.  Even having fixed $\la$ and $\nu$, it is not possible to evaluate the discriminants for general $x$ and $y$.  Hence, we will first fix $x$ (separately choosing $x=0$ and $x=1/2$), and vary $y$ in order to capture the change of behaviour both on the horizon $y=y_h$ and on the $\psi$-axis $y=-1$.

Doing this for both values of $x=0$ and $x=1/2$ , we find that for the first four cases above:
\begin{align}
  \DC{8}  \propto (y+1)^2 (y-y_h)^2  (y-x)^{112},& \quad 
  \DC{9}  \propto (y+1)^4 (y-y_h)^4 (y-x)^{144}, \notag \\
  \DC{10} \propto (y+1)^6 & (y-y_h)^6 (y-x)^{180},
\end{align}
where only the relevant factors have been kept.  
That is, we have only included factors that appear in all the 
three discriminants above. Recall that all three discriminants must 
necessarily vanish for the solution to be type {\bf II} at that point.  We have also neglected factors that vanish for values of $y$ outside its range.

For the final (extremal) case we find for both values of $x=0$ and $x=1/2$ 
that:
\begin{align}
  \DC{8}  \propto (y+1)^2 (y-y_h)^{10}  (y-x)^{112},& \quad
  \DC{9}  \propto (y+1)^4 (y-y_h)^{14} (y-x)^{144}, \notag \\
  \DC{10} \propto (y+1)^6 & (y-y_h)^{20} (y-x)^{180},
\end{align}
and also
\begin{equation}
  \DC{6} \propto (y-y_h)^2 (y-x)^{60}, \quad
  \DC{7} \propto (y-y_h)^6 (y-x)^{84}.
\end{equation}

Now, we fix $y$ depending on the value of $\lambda$ and $\nu$ such that it is outside the horizon. For the first case where the black ring is singly spinning, the ergosurface is at a fixed value of $y$.  For our example, $\la=8/9$, $\nu=1/9$, $y_e=-65/63$.  Thus, we choose three values of $y$ inside the ergoregion ($y=-6/5$), outside the ergoregion ($y=-65/64$) and on the ergosurface ($y_e=-65/63$) and calculate the discriminants to capture their $x-$dependence.  We find that for all three values of $y$
\begin{align}
  \DC{8}  \propto (x-1)^6 (x+1)^6 (y-x)^{112},& \quad
  \DC{9}  \propto (x-1)^{10} (x+1)^{10} (y-x)^{144}, \notag \\
  \DC{10} \propto (x-1)^{16} &(x+1)^{16} (y-x)^{180},
\end{align}
and also
\begin{equation}
  \DC{6} \propto (x-1)^2 (x+1)^2 (y-x)^{60}, \quad
  \DC{7} \propto (x-1)^4 (x+1)^4 (y-x)^{84}.
\end{equation}

For the next four cases for which the black ring is doubly spinning, the ergosurface is not at a fixed value of $y$.  Thus, in each of these cases we pick two values of $y$, one inside the ergoregion and the other outside:
\begin{itemize}
  \item {\bf Case 2} ($\la=7/9$, $\nu=1/9$): $y=-21/20$, $y=-3/2$.
  \item {\bf Case 3} ($\la=8/9$, $\nu=1/9$): $y=-41/40$, $y=-6/5$.
  \item {\bf Case 4} ($\la=1$, $\nu=1/9$): $y=-161/160$, $y=-11/10$.
  \item {\bf Case 5} ($\la=2/3$, $\nu=1/9$): $y=-11/10$, $y=-2$.
\end{itemize}
In all of the cases above, for both values of $y$, we find that
\begin{align}
  \DC{8}  \propto (x-1)^2 (x+1)^2 (y-x)^{112},& \quad 
  \DC{9}  \propto (x-1)^{4} (x+1)^{4} (y-x)^{144}, \notag \\
  \DC{10} \propto (x-1)^{6} &(x+1)^{6} (y-x)^{180}.
\end{align}

In summary, the analysis of the discriminants $\DC{i}$ suggests that the black ring can only be of type {\bf II} or more special at
\begin{equation}
  y=y_h, \quad y=-1, \quad x=\pm 1;
\end{equation}
i.e.\ at the horizon (where we already know it to be type {\bf II} or more special \cite{Lewandowski:2004sh, brwands}), the $\psi$-axis ($y=-1$)  and the $\phi$-axis ($x=\pm 1$).

Finally, we consider the discriminants $\DT{i}$.  To make the analysis easier we use the results obtained from the analysis of the discriminants $\DC{i}$.  Recall that the necessary conditions on $\DT{i}$ for the solution to be type {\bf II} or more special is that:
\begin{equation}
  \DT{2} \geq 0, \quad
  \DT{3} \geq 0, \quad 
  \DT{4} \geq 0, \quad 
  \DT{5} = 0.
\end{equation}
As expected, we find that these conditions are satisfied for $y=y_h$ for all five cases.  However, for the axes ($y=-1$ and $x=\pm 1$) these conditions are only satisfied for the singly spinning black ring.  For the other cases, $\DT{5}=0$.  However, one of the other discriminants may be negative.  Thus, the axes only satisfy the necessary conditions for type {\bf II} or more special for the singly spinning black ring.

Therefore, we conclude that the double black ring is not type {\bf II} or more special except at the horizon.  In addition, for the singly spinning black ring, the axes, $y=-1$ and $x=\pm 1$, also satisfy some of the necessary conditions.

\section{Other examples} 

There are many other higher-dimensional spacetimes of interest.  As a further test of the usefulness of these techniques, we will briefly study what discriminant analysis gives us for a collection of further spacetimes, that look rather different to the asymptotically flat spacetimes considered so far.  In doing so, we hope to build up further intuition for whether discriminant analysis is likely to be a useful tool for studying higher-dimensional spacetimes. 

\subsection{Black holes with nilmanifold horizons}
A 2-parameter family of exact vacuum solutions representing a 5D black hole spacetime with a local 3-dimensional nilmanifold  horizon was found in \cite{CW}.  The metric can be written as
\begin{multline}
  ds^2 = - \left(\frac{2L^2}{11}r^2-\frac M{r^{5/3}}\right)dt^2
         + \frac{dr^2}{\left(\frac{2L^2}{11}r^2-\frac{M}{r^{5/3}}\right)} 
         + r^{4/3}\left(dx^2+dy^2\right) \\
         + r^{8/3}\left(dz-\frac{2L}3xdy\right)^2
\end{multline}
Using the operator $\Tsf$, we can determine the type outside the horizon.  The spacetime admits a spatial isotropy, which the operator $\Tsf$ will inherit; and consequently $\DT{5}=0$.  We can simplify expressions by fixing $r=1$.\footnote{We can do this without loss of generality, as we can recover the $r$-dependence if we note how the scalars and metric scale with $r$.  If $I(L,M)$ is an $n$th order invariant, then we can recover the $r$-dependence as follows: 
\begin{equation*} 
  I(L,M)\mapsto r^{-\tfrac 43n}I(Lr^{2/3},Mr^{-7/3}).
\end{equation*} }
Computing $\DT{4}$, we get
\begin{eqnarray}
  \DT{4} &\propto &(2211 M + 104 L^2 )^2 (-934241 M^2  + 3102 L^2  M + 312 L^4 )^2\\ 
        && \times (432575 M^3  - 61105 L^2  M^2  - 1320 L^4  M + 144 L^6 )^2 \nonumber \\
        && \times (6320919 M^3  - 1484549 L^2  M^2  + 30096 L^4  M + 288 L^6 )^2 \nonumber \\
        && \times (3955732 M^3  - 418055 L^2  M^2  - 27324 L^4  M + 336 L^6 )^2\nonumber \\
        && \times (-251801 M^2  + 3234 L^2  M + 234 L^4 )^2  (-11 M + 2 L^2 )^8.
\end{eqnarray}
When $M=2L^2/11$, the horizon lies at $r=1$, and hence $\DT{4} = 0$.  Away from the horizon this discriminant is generally non-zero (except for a few special values).  The eigenvalue structure is thus $\{1,11(11)\}$, and 
hence the Weyl tensor is of type {\bf  I/G} outside the horizon.  

On the horizon, we get
\begin{equation} 
  \DT{5} = \DT{4} = \DT{3} = 0, \quad 
  \DT{2} \neq 0, \quad 
  F^T_{2}=0; 
\end{equation}
consequently, $\Tsf$ has eigenvalue type $\{(1111)1\}$ on the horizon.  This is consistent with the spacetime being of type {\bf D} on the horizon. 

\subsection{Black holes with solvmanifold horizons}
A black hole with a local 3-dimensional solvmanifold horizon \cite{thurston} was also found in \cite{CW}. 
The metric can be written
\begin{equation}
  ds^2 = -2\left(\frac{L^2}9 r^2-\frac Mr\right)dt^2
          +\frac{dr^2}{2\left(\frac{L^2}9r^2-\frac{M}r\right)}
          +\frac{3}{L^2}\left[r^2(e^{2z}dx^2+e^{-2z}dy^2)+dz^2\right].
\end{equation}
The asymptotics of this spacetime is locally that of a homogeneous Einstein solvmanifold \cite{Nilsol,BH}. Such spaces are generalisations of AdS spacetimes (which is the simplest form of a solvmanifold) and are locally distinct from such. Interestingly, these black hole spacetimes share many of the features of topologial black holes in AdS but are topologically different. 

This has a discrete symmetry $(x,y,z)\mapsto (-x,z,y)$ which interchanges two of the spatial eigenvalues of the operator $\Tsf$.  Consequently, $\DT{5}=0$.  However, computing the discriminant $\DT{4}$ (setting $r=1$) and requiring $M>0$ gives: \footnote{Here, if $I$ is an $n$th order invariant, we can recover the $r$-dependence via 
\begin{equation*} 
  I(L,M)\mapsto r^{-2n}I(Lr,Mr^{-1}).
\end{equation*} } 
\begin{eqnarray}
  \DT{4} &=& (+ve) (L^2 - 9M)^6. 
\end{eqnarray}
Requiring positive mass, i.e.\ $M>0$, this is only zero at the horizon, $L^2=9M$; consequently, this metric  is of type {\bf  I/G} outside of the horizon. 

On the horizon, this degenerates to type $\{(11)(111)\}$. The Weyl operator can easily be computed on the horizon and it is easily verified that it is of type {\bf D}. 

\subsection{5D Generalised Collinson-French spacetimes}
This is not a black hole metric but rather a one-parameter family of cosmological solutions in 5D \cite{SHthesis}.  The family of solutions has the metric:
\begin{multline}
  ds^2 = - dt^2 + t^2 dx^2 + \left(t^{(1-p)^2s}e^{-\sqrt{6}(1+p)rx} dy + \frac{\sqrt{s}}{2r}t dx\right)^2 \\
         + t^{6(1+p)s} e^{4\sqrt{6}rx} dz^2 + t^{6p(1+p)s} e^{4p\sqrt{6}rx} dw^2,
\end{multline}
where 
\begin{equation}
  r = \frac{\sqrt{1+p+p^2}}{5+2p+5p^2},\qquad 
  s = \frac{1}{5+2p+5p^2}.
\end{equation}
For $p=1$, this has an isotropy, so we expect $\DT{5}=0$ here.  To calculate the discriminants we note that this spacetime is self-similar with a time-like homothety.  For an $n$th order invariant, the $t$-dependence will be $t^{-2n}$; therefore, there is no loss of generality to set $t=1$ in the final expressions. In general ($t=1$):
\begin{eqnarray} 
  \DT{5} &\propto& p^{20}(p - 1)^2  (1 + p)^6  (1 + p + p^2 )(57 p^{10} - 81p^9 + 144p^8 + 94p^7 \nn \\ 
        &&  - 121 p^6 + 390 p^5 - 121p^4  + 94p^3  + 144p^2 - 81 p + 57) \nn \\
        && \times (34 p^5  + 17 p^3  + 17 p^2  - 3 p + 7)^2  (7 p^5  - 3 p^4  + 17 p^3  + 17 p^2  + 34)^2. \qquad
\end{eqnarray}
Therefore, so long as $p$ is different from $0$, $\pm 1$ (and some other very special values), the spacetime is of type {\bf I/G}.  

For $p=0$, for example, $\DT{5} = 0$.  Indeed, for $p=0$ \emph{all} invariants are zero; 
hence, it is a VSI$_0$ space \cite{class}.  
{\footnote{ A VSI$_i$ space is a spacetime
for which all of the zeroth to $i$-th order curvature invariants 
(i.e., the scalar invariants constructed from the Riemann tensor and its
first  $i$ covariant derivatives) vanish \cite{class}.}}
Since the case $p=0$ is VSI$_0$, we can consider the {\emph{derivative}} operator:
\begin{equation}
  \tilde{T}^a_{~ c}= C^{b d e f ;a}C_{b d e f; c}
                     -\frac{1}{5} g^a_{~c} C^{b d e f ;h} C_{ b d e f ;h}. 
\end{equation} 
Computing the discriminants in the case $p=0$, we get:
\begin{equation}
{\disc{\tilde{T}}{5}{5}}
= {\disc{\tilde{T}}{5}{4}} = 0, \quad 
{\disc{\tilde{T}}{5}{3}}>0, \quad \text{and}\quad 
{{{}_{\sf\tilde{T}}^{5}\! E_{2}}} =0,
\end{equation}
which shows that the eigenvalue type is $\{(111)11\}$. Therefore, since this metric is VSI$_0$ but not VSI$_1$, this metric is $\mathcal{I}$-non-degenerate \cite{inv}.


\section{Discussion}

The study of higher dimensional black holes is of current interest.  The algebraic classification of
spacetimes has played a crucial role in understanding black holes in 4D, and it is likely to play a
similar role in higher dimensions.  There are a number of algebraic classifications of spacetimes in
higher dimensions (utilizing alignment theory, bivectors and discriminants).  Previous work has produced
a set of necessary (but not sufficient) conditions on the discriminants for a spacetime to be of a
particular (special) algebraic type.  We have demonstrated that this discriminant approach is a reasonably
practical approach in that even in the case of some very complicated metrics it is possible to compute
the relevant discriminants and extract useful information from them.  Indeed, in this paper we have illustrated
these techniques in a useful application to the alignment classification of a number of important
spacetimes.  In particular, we investigated the Sorkin-Gross-Perry soliton, the supersymmetric black
ring, the doubly-spinning black ring, and some other higher dimensional spacetimes.

For the Sorkin-Gross-Perry (SGP) soliton spacetimes all of $\DC{8}$, $\DC{9}$, and $\DC{10}$
vanish.  In the general case we have that all other $\DC{i}$ are non-zero.  Note that even with
$ \DC{8}=\DC{9}=\DC{10}= 0$, this eigenvalue structure implies type {\bf{G}} or
{\bf{I}} because the SGP spacetime has a spatial isotropy.  These SGP examples thus illustrate the 
application of Proposition 2.5 in
the discrimination analysis.  Note that on $R=R_0$, $\DC{4} = \DC{5} = \DC{6} = \DC{7} = 0$, and so on
this line the spacetime is of type {\bf D}.

We next considered the 5D supersymmetric rotating black (SBR) holes (that include the extremal charged
rotating BMPV black hole of \cite{BMPV}).  Since $ \DC{8}=\DC{9}=\DC{10}= 0$, this is a
signal that it is of type {\bf II/D} on the horizon.  Explicitly, the eigenvalue type of the Weyl
bivector operator is $\{1(111)(111111)\}$, while the operator $T_a^{~b}=C_{cdea}C^{cdeb}$ is of eigenvalue
type:  $\{(11)(111)\}$, on the horizon.  As long as $\DT{5}$ is non-zero, the Weyl tensor is of type
{\bf I/G}.

As a demonstration of the utility of the discriminant techniques, we also proved that the doubly
spinning black ring (DSBR) is not of type {\bf{II}} or more special, except on the horizon.  Indeed, we
first showed, using the discriminants $\DC{i}$, that the black ring can only be of type {\bf II} or
more special at the horizon and on the $\psi$-axis and the $\phi$-axis.  We then considered the
discriminants $\DT{i}$, and showed that the necessary conditions for the solution to be of type {\bf
II} or more special on these axes are only satisfied for the singly spinning black ring.  We concluded
that the double black ring is only of type {\bf II} or more special at the horizon.

There are many other higher dimensional spacetimes of interest.  We also analysed the {black hole
solution with a {\sf Sol}-horizon } and the {black hole solution with a {\sf Nil}-horizon} and the {5D
generalised Collinson-French} solution.

\section*{Acknowledgements}  
 
The work was supported by NSERC of Canada (AC), by EPSRC
of the UK (MG) and by a Leiv Erikson mobility grant from
the Research Council of Norway, project no:  {\bf 200910/V11} (SH).  
MND and MG thank AC and Dalhousie University
for hospitality during their visit in October 2010, when this
work was initiated.

\appendix

\section{Discriminant Analysis: The Algorithm}

At a point in spacetime, any curvature operator $\Rsf$ is a linear map from one vector space to another.
In this Appendix we describe an
algorithm for determining  
the eigenvalue structure of this linear operator
and, in particular, give criteria for different Segre types.  This is a
review of the algorithm described in \cite{DISCRIM}, based on the work of
\cite{Yang:1996,Yang:1997,Liu}, to analyse the characteristic polynomial of the operator.

The criteria for various eigenvalue types will be given in terms of a set of `syzygies' 
(scalar polynomial invariants) which can be used to characterise the various eigenvalue cases; i.e., 
they are \emph{discriminants}.  For our purposes, for an operator $\Rsf$ acting on a $n$-dimensional 
vector space, the relevant discriminants will be the ten spacetime scalars $\disc{R}{n}{i}$, $1\leq i \leq n$.

\subsection{Discriminant sequences and primary syzygies}
More explicitly, we can write the characteristic equation as a polynomial
\begin{equation} \label{poly}
f(\lambda) \equiv \det(\lambda {\sf 1}-\Rsf )
           \equiv a_0\lambda^n+a_1\lambda^{n-1}+\dots a_i\lambda^{n-i}+\ldots +a_n 
           = 0.
\end{equation}
Choosing a normalisation such that $a_0=1$, the other coefficients $a_i$ can be expressed explictely in terms of the polynomial invariants 
\begin{equation}
  R_1 \equiv \Tr(\Rsf),  \quad
  R_2 \equiv \Tr(\Rsf^2), \quad 
  R_3 \equiv \Tr(\Rsf^3), \quad\text{etc},  
\end{equation}
of $\Rsf$, by\footnote{Note that if $a_n=0$, then the eigenvalue equation trivially 
factorises and we have a zero eigenvalue.  Furthermore, if $a_n = a_{n-1} = \ldots = a_{n-k} = 0$, 
then there exists a zero eigenvalue of multiplicity $k+1$.  Therefore, before beginning 
the algorithm one should check for the existence of zero eigenvalues, and work with the reduced polynomial $f(\la)/\la^k$.}
\begin{equation}\label{defai}
  a_{i}=\frac{(-1)^i}{i!}\det 
        \begin{bmatrix} 
          R_1 & 1 & 0 & \cdots & 0 \\ 
          R_2 & R_1 & 2 & \ddots & \vdots \\ 
          R_3 & R_2 & R_1 & \ddots & 0 \\ 
          \vdots & \ddots & \ddots & \ddots & (i-1) \\ 
          R_i & \dots & R_3 & R_2 & R_1  
        \end{bmatrix}.
\end{equation}

Next, define the $(2n+1)\times(2n+1)$ discrimination matrix 
\begin{equation} 
\mathrm{Disc}(f) = 
\begin{bmatrix} 
  a_0    & a_1    & a_2      & \cdots & a_n     & 0       & \cdots & 0       & 0\\ 
  0      & na_0   & (n-1)a_1 & \cdots & a_{n-1} & 0       & \cdots & 0       & 0\\  
  0      & a_0    & a_1      & \cdots & a_{n-1} &a_n      &        & 0       & 0\\ 
  0      & 0      &na_0      & \cdots & 2a_{n-2}& a_{n-1} &        & 0       & 0\\ 
  \vdots & \vdots &          &        & \vdots  & \vdots  &        & \vdots  & \vdots \\ 
  0      & 0      & \cdots   & 0      & na_0    & (n-1)a_1&\cdots  & a_{n-1} & 0\\ 
  0      & 0      & \cdots   & 0      & a_0     & a_1     & \cdots & a_{n-1} & a_n 
\end{bmatrix} ,
\end{equation}
or more explicitely
\begin{eqnarray}
  \mathrm{Disc}(f)_{2i+1,j} 
    &=& \left\{ \begin{array}{ll}
                0,         & j   \leq i \\
                a_{j-i-1}, & i+1 \leq j \leq i+1+n \\
                0,         & n+i+2 \leq j \leq 2n+1
              \end{array}\right. ,\\
  \mathrm{Disc}(f)_{2i,j} 
     &=& \left\{ \begin{array}{ll}
                0,                & j   \leq i \\
                (n-j+1)a_{j-i-1}, & i+1 \leq j \leq i+1+n \\
                0,                & n+i+2 \leq j \leq 2n+1
              \end{array}\right.
\end{eqnarray}

We are now in a position to make the main definition of this Appendix:
\begin{defn}
  Let $\Rsf$ be an operator with characteristic polynomial $f$, mapping an $n$-dimensional vector 
  space to itself, as above.  Then, $\disc{R}{n}{i}$ is defined to be the determinant of the submatrix formed from the first $2i$ rows and $2i$ columns of $\mathrm{Disc}(f)$.
  
  The \emph{discriminant sequence} of the polynomial $f$ is the sequence
  \begin{equation} \label{discseq}
    \left[\disc{R}{n}{1},\disc{R}{n}{2},\ldots,\disc{R}{n}{n}\right].
  \end{equation}             
When rewritten in terms of the curvature invariants $\{R_1, R_2, \ldots \}$, using (\ref{defai}), these discriminants are the \emph{primary syzygies} $\disc{R}{n}{i}$ for the operator $\Rsf$.  
\end{defn}

Note that if $\disc{R}{n}{i}=0$ then there can be at most $(n-1)$ distinct eigenvalues, and hence there must be a repeated eigenvalue.  More generally, if $\disc{R}{n}{n-j} =\ldots = \disc{R}{n}{n} = 0$, then there are at most $n-j$ distinct eigenvalues.  We can use this to find necessary conditions for the an operator to have a particular algebraic type in the alignment classification. 
In particular, this can be used to determine the number of real roots of the characteristic equation, as described in Section \ref{sec:disc}.

\paragraph{Multiple factor sequence.}  
For polynomials of order 3 or less, the discriminants $\disc{R}{n}{i}$ are sufficient to determine the complete eigenvalue structure.  However, for higher-order polynomials (corresponding to operators acting on higher-dimensional vector spaces), we need more information.

To get this, consider the following definition:
\begin{defn}
  Let $M(k,l)$ be the submatrix of the discriminant matrix $Disc(f)$ formed by the first $2k$ rows and the first $(2k-1)$ columns + $(2k+l)$th column and construct the polynomials: 
  \begin{equation} 
    \Delta_k(f) = \sum_{i=0}^k \det[M(n-k,i)] x^{k-i}, 
  \end{equation} 
  for $k=0,1,...,n-1$.  The \emph{multiple factor sequence} of the polynomial $f$ is
  \begin{equation}
    \left[\Delta_0(f),\Delta_1(f),...,\Delta_{n-1}(f)\right]
  \end{equation}
\end{defn}
 
This sequence is useful because
\begin{lem}[\cite{Yang:1996,Yang:1997}]
  If the number of zeros in the revised sign list of the discriminant sequence of $f(x)$ is $k$, then the greatest common divisor of $f$ and its derivative $f'$ is given by $\Delta_k(f)={\rm g.c.d.}(f(x),f'(x))$.  Hence, the g.c.d.\ of $f$ and $f'$ is thus always in the multiple factor sequence.
\end{lem}  
Indeed, the polynomial $\Delta(f)(x)\equiv {\rm g.c.d.}(f(x),f'(x))$ is the repeated part of $f(x)$, because if $\Delta (f)$ has $k$ real roots of multiplicities $n_1$, $n_2$, ...,$n_k$, and $f$ has $m$ distinct real roots, then $f$ has $k$  real roots of multiplicities $n_1+1$, $n_2+1$, ...,$n_k+1$, and $m-k$ simple real roots (similarly for complex roots). Therefore, by considering $\Delta(f)$ we reduce the multiplicities of all the roots by 1.  
 
We can now consider the discriminants of the polynomial $\Delta(f)$ in the same way as we computed the
discriminant sequence of $f$.  We will call the discriminant sequence of $\Delta(f)$
$\{{}^nE_1,{}^nE_2,{}^nE_3,...,{}^nE_k\}$.  We can now use these to determine the revised sign list of
the $E$-sequence, etc.  We can repeat this procedure and consider $\Delta(\Delta(f))\equiv\Delta^2(f)$,
$\Delta^3(f)$ etc (we label the next discriminant sequence $F$ in the table below).

Iterating this, we eventually arrive at some $j$ for which the revised sign list of $\Delta^j(f)$ contains no zeros.  We can then  compute the number of real/complex distinct roots of $\Delta^j(f)$.  This in turn allows us to determine the roots and multiplicities of $\Delta^{j-1}(f)$, which again enables us to determine the roots and multiplicites of $\Delta^{j-2}(f)$ etc.  At the end of this process, we have a complete root classification for $f(x)$.  Further details of this procedure are described in \cite{Yang:1996,Yang:1997,inv}. 

Using these methods, it is possible to give necessary and sufficient conditions on the various discriminants for
an operator to have a particular eigenvalue type.  From these conditions, we can extract the necessary
conditions on $\DC{8}$, $\DC{9}$, $\DC{10}$ etc that we make extensive use of in this paper.

For the 10-dimensional trace-free Weyl operator, a partial table demonstrating this is as follows:
\begin{center}
  \begin{tabular}{|c|c|c|c|c|c|c|c|}
  \hline 
 $\DC{10}$ & $\DC{9}$ & $\DC{8}$ &  $\DC{7}$ & ${}^{10}E_2$ 
                      & ${}^{10}F_2$ & ${}^{10}F_3$& Eigenvalue type \\
 \hline
 $\neq 0$ &  &  &  & & & & $\{111..1\}$ \\
 0 & $\neq 0 $ & &  &   & & &$\{(11)11...1\}$ \\
 0 & 0 & $\neq 0$  &   & $\neq 0$ & & & $\{(11)(11)1...1\}$\\ 
 0 & 0 & $\neq 0$  &   & 0  & & & $\{(111)1...1\}$ \\
 0 & 0 & 0  & $\neq 0$  &  &   &$\neq 0$ &$\{(11)(11)(11)1..1\}$ \\
 0 & 0 & 0  & $\neq 0$  &  &$\neq 0$ &  0 & $\{(111)(11)1..1\}$ \\
 0 & 0 & 0  & $\neq 0$  & & 0       &   0 & $\{(1111)11..1\}$ \\
 \hline
 \end{tabular}\\
\end{center}


\providecommand{\href}[2]{#2}
\begingroup\raggedright

\endgroup

\end{document}